\documentstyle[preprint,aps]{revtex}
\tightenlines

\begin{document}

\draft
\title
{Fractal statistics, fractal index and fractons}

\author
{\bf WELLINGTON DA CRUZ\footnote{E-mail: wdacruz@exatas.uel.br}}

\address
{Departamento de F\'{\i}sica,\\
 Universidade Estadual de Londrina, Caixa Postal 6001,\\
Cep 86051-970 Londrina, PR, Brazil\\}
 
\date{\today}

\maketitle

\begin{abstract}

The concept of fractal index is introduced in connection with 
the idea of universal 
class $h$ of particles or quasiparticles, termed fractons, which obey 
fractal statistics. We show the relation between fractons and 
conformal field theory(CFT)-quasiparticles taking into 
account the central charge $c[\nu]$ and the particle-hole duality 
$\nu\longleftrightarrow\frac{1}{\nu}$, for integer-value $\nu$ of the 
statistical parameter. The Hausdorff dimension $h$ which 
labelled the universal classes of 
particles and the conformal anomaly are therefore related. 
We also establish a connection between Rogers 
dilogarithm function, Farey series of rational numbers and the 
Hausdorff dimension.

\end{abstract}


We consider the conformal field theory(CFT)-quasiparticles 
( edge excitations ) in connection with the concept 
of fractons introduced in\cite{R1}. These excitations have been considered at 
the edge of the quantum Hall systems which in the fractional regime assume the 
form of a chiral Luttinger liquid\cite{R2}. Beyond this, 
conformal field theories have been exploited in a variety of contexts, 
including statistical mechanics at the critical 
point, field theories, string theory, and in various 
branches of mathematics\cite{R3}.

In this Letter, we suppose that the fractal statistics obeyed by 
fractons are shared by CFT-quasiparticles. Thus, the central charge, a 
model dependent constant is related to the universal class $h$ of the 
fractons. We define the {\it fractal index} associated with 
these classes as 

\begin{equation}
\label{e.1}
i_{f}[h]=\frac{6}{\pi^2}\int_{\infty(T=0)}^{1(T=\infty)}
\frac{d\xi}{\xi}\ln\left\{\Theta[\cal{Y}(\xi)]\right\}
\end{equation}

\noindent where

\begin{eqnarray}
\Theta[{\cal{Y}}]=
\frac{{\cal{Y}}[\xi]-2}{{\cal{Y}}[\xi]-1}
\end{eqnarray}

\noindent is the single-particle partition function of the 
universal class $h$ and $\xi=\exp\left\{(\epsilon-\mu)/KT\right\}$, 
has the usual definition. The function ${\cal{Y}}[\xi]$ satisfies 
the equation 

\begin{eqnarray}
\label{e.46} 
\xi=\left\{{\cal{Y}}[\xi]-1\right\}^{h-1}
\left\{{\cal{Y}}[\xi]-2\right\}^{2-h}.
\end{eqnarray}

We note that the general solution of the algebraic 
equation derived from this last one is of the form

\[ 
{\cal{Y}}_{h}[\xi]=f[\xi]+{\tilde{h}}
\]

or

\[ 
{\cal{Y}}_{\tilde{h}}[\xi]=g[\xi]+h,
\]

\noindent where ${\tilde{h}}=3-h$, is a 
{\it duality symmetry}\footnote{This means that fermions($h=1$) 
and bosons($h=2$) are dual objects. As a result we have a {\it 
fractal supersymmetry}, since for the particle with spin $s$ 
within the class $h$, its dual $s+\frac{1}{2}$ is within the class ${\tilde{h}}$.} 
between 
the classes. The functions $f[\xi]$  and $g[\xi]$ at least 
for third, fourth degrees algebraic equation differ by plus and minus signs 
in some terms of their expressions.

The particles within each class $h$ satisfy 
specific {\bf fractal statistics}\footnote{In fact, we have here 
{\it fractal functions}\cite{R4}. }\cite{R1}

\begin{eqnarray}
\label{e.h} 
n&=&\xi\frac{\partial}{\partial{\xi}}\ln\Theta[{\cal{Y}}]\nonumber\\
&=&\frac{1}{{\cal{Y}}[\xi]-h}
\end{eqnarray}

\noindent and the fractal parameter
\footnote{This parameter describes the properties  of the path
({\it fractal curve}) of the quantum-mechanical particle.}  
(or Hausdorff dimension) 
$h$ defined in the interval $1$$\;$$ < $$\;$$h$$\;$$ <$$\;$$ 2$ 
is related to the  spin-statistics relation $\nu=2s$ 
through the {\it fractal spectrum}

\begin{eqnarray}
\label{e.7}
&&h-1=1-\nu,\;\;\;\; 0 < \nu < 1;\;\;\;\;\;\;\;\;
 h-1=\nu-1,\;
\;\;\;\;\;\; 1 <\nu < 2;\\
&&etc.\nonumber
\end{eqnarray}

\noindent For $h=1$ we have fermions, with ${\cal{Y}}[\xi]=\xi+2$, 
$\Theta[1]=\frac{\xi}{\xi+1}$ and $i_{f}[1]=
\frac{6}{\pi^2}\int_{\infty}^{1}\frac{d\xi}{\xi}
\ln\left\{\frac{\xi}{\xi+1}\right\}=\frac{1}{2}$. 
For $h=2$ we have bosons, with  
${\cal{Y}}[\xi]=\xi+1$, $\Theta[2]=
\frac{\xi-1}{\xi}$ and $i_{f}[2]=\frac{6}{\pi^2}
\int_{\infty}^{1}\frac{d\xi}{\xi}
\ln\left\{\frac{\xi-1}{\xi}\right\}=1$. On the other hand, 
for the universal class 
$h=\frac{3}{2}$, 
we have fractons with ${\cal{Y}}[\xi]=\frac{3}{2}+\sqrt{\frac{1}{4}+\xi^2}$, 
$\Theta\left[\frac{3}{2}\right]=\frac{\sqrt{1+4\xi^2}-1}{\sqrt{1+4\xi^2}+1}$ and  
$i_{f}\left[\frac{3}{2}\right]=\frac{6}{\pi^2}\int_{\infty}^{1}\frac{d\xi}{\xi}
\ln\left\{\frac{\sqrt{1+4\xi^2}-1}{\sqrt{1+4\xi^2}+1}\right\}=\frac{3}{5}$.

The distribution function for each class $h$ above are given by

\begin{eqnarray}
n[1]&=&\frac{1}{\xi+1},\\
n[2]&=&\frac{1}{\xi-1},\\
n\left[\frac{3}{2}\right]&=&\frac{1}{\sqrt{\frac{1}{4}+\xi^2}},
\end{eqnarray}

\noindent i.e. we have the Fermi-Dirac distribution, the Bose-Einstein 
distribution and the fracton distribution of the universal 
class $h=\frac{3}{2}$, 
respectively . Thus, our formulation generalizes {\it in a natural 
way} the fermionic and 
bosonic distributions for particles assuming 
rational or irrational values for the spin quantum number $s$. In this way, 
our approach can be understood as a {\it quantum-geometrical} description of 
the statistical laws of Nature. This means that the (Eq.\ref{e.h}) 
captures the observation about the fractal characteristic of the 
{\it quantum-mechanical} path, which reflects the 
Heisenberg uncertainty principle. 

The fractal index as defined has a connection with the central 
charge or conformal anomaly $c[\nu]$, a dimensionless number which 
characterizes conformal field theories in two dimensions. This way, 
we verify that 
the conformal anomaly is associated with universality classes, 
i.e. universal classes $h$ 
of particles. Now, we consider the particle-hole duality 
$\nu\longleftrightarrow\frac{1}{\nu}$ for integer-value 
$\nu$ of the statistical parameter in connection 
with the universal class $h$. For bosons and fermions, we have

\[
\left\{0,2,4,6,\cdots\right\}_
{h=2}
\]

and

\[
\left\{1,3,5,7,\cdots\right\}_
{h=1}
\]

such that, the central charge for $\nu$ {\it even} is defined by

\begin{eqnarray}
\label{e.11}
c[\nu]=i_{f}[h,\nu]-i_{f}\left[h,\frac{1}{\nu}\right]
\end{eqnarray}

\noindent and for $\nu$ {\it odd} is defined by

\begin{eqnarray}
\label{e.12}
c[\nu]=2\times i_{f}[h,\nu]-i_{f}\left[h,\frac{1}{\nu}\right],
\end{eqnarray}

\noindent where $i_{f}[h,\nu]$ means the fractal 
index of the universal class $h$ which contains the particles 
with distinct spin values which obey specific 
fractal statistics. We assume that the fractal 
index $i_{f}[h,\infty]=0$ and we obtain, for example, the results

\begin{eqnarray}
&&c[0]=i_{f}[2,0]-i_{f}[h,\infty]=1;\nonumber\\
&&c[1]=2\times i_{f}[1,1]-i_{f}[1,1]=\frac{1}{2};\nonumber\\
&&c[2]=i_{f}[2,2]-i_{f}\left[\frac{3}{2},\frac{1}{2}\right]=
1-\frac{3}{5}=\frac{2}{5};\\
&&c[3]=2\times i_{f}[1,3]-i_{f}\left[\frac{5}{3},\frac{1}{3}\right]=
1-0.656=0.344;\nonumber\\
&&etc,\nonumber
\end{eqnarray}

\noindent where the fractal index for $h=\frac{5}{3}$ is obtained from 

\begin{eqnarray}
&&i_{f}\left[\frac{5}{3}\right]=\frac{6}{\pi^2}\int_{\infty}^{1}\frac{d\xi}{\xi}
\\
&&\times\ln\left\{\frac{{\sqrt[3]{\frac{1}{27}+\frac{\xi^3}{2}
+\frac{1}{18}\sqrt{12\xi^3+81\xi^6}}}+
\frac{1}{9\sqrt[3]{\frac{1}{27}+\frac{\xi^3}{2}+\frac{1}{18}
\sqrt{12\xi^3+81\xi^6}}}-\frac{2}{3}}
{{\sqrt[3]{\frac{1}{27}+\frac{\xi^3}{2}
+\frac{1}{18}\sqrt{12\xi^3+81\xi^6}}}+
\frac{1}{9\sqrt[3]{\frac{1}{27}+\frac{\xi^3}{2}+\frac{1}{18}
\sqrt{12\xi^3+81\xi^6}}}+\frac{1}{3}}\right\}\nonumber\\
&&=0.656\nonumber
\end{eqnarray}

\noindent and for its dual we have

\begin{eqnarray}
&&i_{f}\left[\frac{4}{3}\right]=\frac{6}{\pi^2}\int_{\infty}^{1}\frac{d\xi}{\xi}
\\
&&\times\ln\left\{\frac{{\sqrt[3]{-\frac{1}{27}+\frac{\xi^3}{2}
+\frac{1}{18}\sqrt{-12\xi^3+81\xi^6}}}+
\frac{1}{9\sqrt[3]{-\frac{1}{27}+\frac{\xi^3}{2}+\frac{1}{18}
\sqrt{-12\xi^3+81\xi^6}}}-\frac{1}{3}}
{{\sqrt[3]{-\frac{1}{27}+\frac{\xi^3}{2}
+\frac{1}{18}\sqrt{-12\xi^3+81\xi^6}}}+
\frac{1}{9\sqrt[3]{-\frac{1}{27}+\frac{\xi^3}{2}+\frac{1}{18}
\sqrt{-12\xi^3+81\xi^6}}}+\frac{2}{3}}\right\}\nonumber\\
&&=0.56.\nonumber
\end{eqnarray}

\noindent The correlation 
between the classes $h$ of particles and their fractal index, show us 
{\it a robust} consistence in accordance with the unitary
$c[\nu]$$\;$$ <$$\;$$ 1$ representations\cite{R3}. Therefore, 
since $h$ is defined within the interval 
$ 1$$\;$$ < $$\;$$h$$\;$$ <$$\;$$ 2$, the corresponding fractal index 
is into the interval $0.5$$\;$$ < $$\;$$i_{f}[h]$$\;$$ <$$\;$$ 1$. Howewer, 
the central charge $c[\nu]$ can assumes values less than $0.5$. Thus, we 
distinguish two concepts of central charge, one is related to the 
universal classes $h$ and the other is related 
to the particles which belong to these classes. 

For the statistical 
parameter in the interval $0$$\;$$ < $$\;$$\nu$$\;$$ <$$\;$$ 1$ (the first 
elements of each class $h$), $c[\nu]
=i_{f}[h,\nu]$, as otherwise we obtain different values. In another way, 
the central charge $c[\nu]$ can be obtained using the 
Rogers dilogarithm function\cite{R6}, i.e.

\begin{equation}
\label{e.16}
c[\nu]=\frac{L[x^{\nu}]}{L[1]},
\end{equation}

\noindent with $x^{\nu}=1-x$,$\;$ $\nu=0,1,2,3,etc.$ and 

\begin{equation}
L[x]=-\frac{1}{2}\int_{0}^{x}\left\{\frac{\ln(1-y)}{y}
+\frac{\ln y}{1-y}\right\}dy,\; 0 < x < 1.
\end{equation}

This way, we observe that our formulation to 
the universal class $h$ of particles with any values of spin $s$ 
establishes a connection between Hausdorff dimension $h$ and 
the central charge $c[\nu]$, in a manner unsuspected till now. 
Besides this, we have obtained a connection between $h$ and the 
Rogers dilogarithm function, through the fractal index defined 
in terms of the partition function associated with the universal 
class $h$ of particles. Thus, considering 
the Eqs.(\ref{e.11}, \ref{e.12}) and the Eq.(\ref{e.16}), we have

\begin{eqnarray}
\frac{L[x^{\nu}]}{L[1]}&=&
i_{f}[h,\nu]-i_{f}\left[h,\frac{1}{\nu}\right],\; 
\nu=0,2,4,etc.\\
\frac{L[x^{\nu}]}{L[1]}&=&
2\times i_{f}[h,\nu]-i_{f}\left[h,\frac{1}{\nu}\right],\;
\nu=1,3,5,etc.
\end{eqnarray}

 Also in\cite{R1} we have established 
a connection between the fractal parameter $h$ and the Farey 
series of rational numbers, therefore once the classes $h$ satisfy all the 
properties of these series we have an infinity 
collection of them. In this sense, we clearly establish a connection 
between number theory and the Rogers dilogarithm function. Given that 
the fractal parameter is an irreducible number $h=\frac{p}{q}$, 
the classes satisfy the properties\cite{R7}

P1. If $h_{1}=\frac{p_{1}}{q_{1}}$ and 
$h_{2}=\frac{p_{2}}{q_{2}}$ are two consecutive fractions 
$\frac{p_{1}}{q_{1}}$$ >$$ \frac{p_{2}}{q_{2}}$, then 
$|p_{2}q_{1}-q_{2}p_{1}|=1$.

P2. If $\frac{p_{1}}{q_{1}}$, $\frac{p_{2}}{q_{2}}$,
$\frac{p_{3}}{q_{3}}$ are three consecutive fractions 
$\frac{p_{1}}{q_{1}}$$ >$$ \frac{p_{2}}{q_{2}} 
$$>$$ \frac{p_{3}}{q_{3}}$, then 
$\frac{p_{2}}{q_{2}}=\frac{p_{1}+p_{3}}{q_{1}+q_{3}}$.

P3. If $\frac{p_{1}}{q_{1}}$ and $\frac{p_{2}}{q_{2}}$ are 
consecutive fractions in the same sequence, then among 
all fractions\\
 between the two, 
$\frac{p_{1}+p_{2}}{q_{1}+q_{2}}$
 is the unique reduced
fraction with the smallest denominator.

For example, consider the Farey series of order 6, denoted by 
the $\nu$ sequence

\begin{eqnarray}
(h,\nu)&=&\left(\frac{11}{6},\frac{1}{6}\right)\rightarrow 
\left(\frac{9}{5},\frac{1}{5}
\right)\rightarrow
\left(\frac{7}{4},\frac{1}{4}\right)\rightarrow
 \left(\frac{5}{3},\frac{1}{3}\right)\rightarrow\nonumber\\
&&\left(\frac{8}{5},\frac{2}{5}\right)\rightarrow 
\left(\frac{3}{2},\frac{1}{2}\right)
\rightarrow \left(\frac{7}{5},\frac{3}{5}\right)
\rightarrow 
\left(\frac{4}{3},\frac{2}{3}\right) \rightarrow\\ 
&&\left(\frac{5}{4},\frac{3}{4}\right) \rightarrow 
\left(\frac{6}{5},\frac{4}{5}\right) \rightarrow 
\left(\frac{7}{6},\frac{5}{6}\right) \rightarrow 
\cdots.\nonumber
\end{eqnarray}

\noindent Using the fractal spectrum ( Eq.\ref{e.7} ), we can obtain other 
sequences which satisfy the Farey properties and for the classes
 
\[
h=\frac{11}{6},\frac{9}{5},
\frac{7}{4},\frac{5}{3},
\frac{8}{5},\frac{3}{2},
\frac{7}{5},\frac{4}{3},
\frac{5}{4},\frac{6}{5},
\frac{7}{6},\cdots,
\] 

and ( note that these ones are dual classes, ${\tilde{h}}=3-h$ ) 
we can calculate the fractal 
index taking into account the Rogers dilogarithm function or the 
partition function associated with each $h$.

In summary, we have obtained a connection 
between fractons and CFT-quasiparticles. This was implemented with 
the notion of the {\it fractal index} associated with 
the universal class $h$ of the fractons. This way, 
fractons and CFT-quasiparticles satisfy a specific 
fractal statistics. A connection between Rogers 
dilogarithm function, Farey series of rational numbers and 
Hausdorff dimension $h$, also was established. The idea of fractons as 
quasiparticles has been 
explored in the contexts of the fractional quantum Hall effect\cite{R1}, 
high-$T_{c}$ superconductivity\cite{R8} and Luttinger liquids\cite{R9}. A 
connection between fractal statistics and black hole entropy also was 
exploited in\cite{R10}. Finally, a fractal-deformed Heisenberg algebra for 
each class of fractons was introduced in\cite{R11}.

\end{document}